\documentclass[a4paper,12pt]{iopart}

\usepackage{graphicx}
\usepackage{dcolumn}
\usepackage{bm}
\usepackage{hyperref}
\usepackage{color}
\usepackage{ulem}
\usepackage{iopams}
\usepackage{setstack} 

\begin{document}

\title[
Local photo-excitation of shift current in noncentrosymmetric systems
]{
Local photo-excitation of shift current in noncentrosymmetric systems
}

\author{Hiroaki Ishizuka$^1$ and Naoto Nagaosa$^{1,2}$}
\address{
$^1$Department of Applied Physics, The University of Tokyo, Bunkyo, Tokyo, 113-8656, JAPAN 
}
\address{
$^2$RIKEN Center for Emergent Matter Sciences (CEMS), Wako, Saitama, 351-0198, JAPAN
}
\ead{ishizuka@appi.t.u-tokyo.ac.jp}

\begin{abstract}
  Photocurrent in solids is an important phenomenon with many applications including the solar cells. In conventional photoconductors, the electrons and holes created by light irradiation are separated by the external electric field, resulting in a current flowing into electrodes. Shift current in noncentrosymmetric systems is distinct from this conventional photocurrent in the sense that no external electric field is needed, and, more remarkably, is driven by the Berry phase inherent to the Bloch wavefunction. It is analogous to the polarization current in the ground state but is a d.c. current continuously supported by the nonequilibrium steady state under the pumping by light.  Here we show theoretically, by employing Keldysh-Floquet formalism applied to a simple one-dimensional model, that the local photo excitation can induce the shift current which is independent of the position and width of the excited region and also the length of the system.  This feature is in stark contrast to the conventional photocurrent, which is suppressed when the sample is excited locally at the middle and increases towards the electrodes. This finding reveals the unconventional nature of shift current and will pave a way to design a highly efficient photovoltaic effect in solids.
\end{abstract}

\vspace{2pc}

\section{Introduction}

Photo-carrier generation and the associated photocurrent is a subject of intensive studies in condensed matter physics and its applications. For example, the excitation spectrum of photocurrent together with the optical absorption spectrum offers a powerful tool to investigate the electronic states in bulk solids. Namely, the absorption can occur both by the excitons, i.e., the bound states of electron and hole, and electron-hole continuum, while the photocurrent can be carried only by the latter, as the exciton is a charge neutral object and cannot contribute to the current.  By increasing the external electric field for the photocurrent, the excitons split into electrons and holes and begins to contribute to the photocurrent. In solar cells, the external electric field is generated by the p-n junction, where the electrons and holes feel the potential gradient in the opposite directions and are separated~\cite{Nelson2003,Wuerfel2009}. 

In contrast to the conventional view of the photocurrent discussed above, the shift current in noncentrosymmetric systems has been attracting intensive interest~\cite{Fridkin2001,Choi2009,Young2012_1,Young2012_2,Grinberg2013,Nie2015,Shi2015,deQuilettes2015,Morimoto2016}. In this phenomenon, since the system itself has a fixed ``direction'', the light irradiation alone can produce the d.c. current without the external electric field.  One might think that the internal electric field plays a similar role to that of the external one. However, the situation is much more subtle.  Due to the periodicity of the crystal potential, there is no macroscopic potential gradient. Microscopically, the band structure has certain symmetry between $\vec{k}$ and $-\vec{k}$ due to the time-reversal symmetry $T$ ( $\vec{k}$ is the crystal momentum ).  Namely, $T$-symmetry gives the relation $\varepsilon_{\uparrow} (\vec{k}) = \varepsilon_{\downarrow}(-\vec{k})$ with $\uparrow(\downarrow)$ represents the up (down) spin.  Therefore, it is highly nontrivial whether the excitation of electrons and holes contributes to the shift current.  The lack of inversion symmetry $I$, however, opens the possibility of the Berry phase of Bloch wavefunction to be finite. (More precisely, one can choose the gauge of the Bloch wavefunction where the Berry phase connection is zero in the presence of both $T$ and $I$ symmetries, i.e., no Berry phase contribution to material properties.) It have been shown that, the shift current comes from the Berry connection of the Bloch wavefunction $|\vec{k}\,n\rangle$ for conduction ($n=c$) and valence $(n=v$) bands~\cite{Kral2000,Sipe2000,Morimoto2016}. An important observation here is that the Berry connection has the meaning of the intracell coordinate in the sense that the position $\vec{r}_n$ of the wavepacket made from the Bloch wavefunctions shifts proportionally by the Berry connection. Indeed, it is related to the polarization in ferroelectric materials~\cite{Resta1992,King-Smith1993}.

This scenario for the ground state finds its generalization to the nonequilibrium steady state, where the continuous pump from the valence to conduction band occurs by the photo-excitation. In this case, the {\it difference} of the Berry phase between the conduction and valence bands corresponds to the shift of the electron position induced by the transition, and contribute to the d.c. photocurrent. This is the picture of shift current, and hence is different from the transport current supported by the free carriers. It is more quantum mechanical origin, i.e., cannot be attributed to the motion of individual carriers, and is expected to be nonlocal in nature. Therefore, the optical response of noncentrosymmetric photovoltaic systems to local photo-excitation is nontrivial.

To study the shift current due to the local photo-excitation, in this paper, we consider a noncentrosymmetric non-interacting fermion chain as a basic model that exhibits shift current. As we ignore the Coulomb interaction of electrons, our model has no internal bias due to the ferroelectric polarization, therefore, no photocurrent from the conventional mechanism similar to photoconductors. We study the model by a nonequilibrium Green's function technique used to study nonequilibrium phenomena in semiconductors~\cite{Haug2008} and in organic solids~\cite{Wei2006,Wei2007,Wei2007_2}; the details of the method are elaborated in appendices. By applying the method, we numerically study the shift current induced by the local photo-excitation using the fermion chain up to $N=1200$ sites. Although there are no potential gradient, we find that a finite photocurrent appears in this model. Furthermore, we find that the locally excited shift current in the fermion chain behaves in a distinct manner from the photocurrent in photoconductors and in p-n junctions, as shown in Fig.~\ref{fig:bdchain}. In particular, we find that the magnitude of shift current does not depend on the position of the photoexcitation. This is a contrasting feature to the conventional case, in which the current is suppressed for the photoexcitation at the center of the chain. This is an interesting feature that experimentally distinguishes the shift current from the conventional photocurrent. Furthermore, this is an advantageous feature for application to highly efficient optoelectronic devices.

We also find that the magnitude of photocurrent does not change by increasing the width of the bonds excited. As presented in Sec.~\ref{sec:results:lightwidth}, we discuss that this is likely to be a consequence of high carrier density excited by light.

The remainder of the paper is organized as follows. In Sec.~\ref{sec:model}, we introduce the model we consider in this paper. In Sec.~\ref{sec:results}, we present the result of theoretical calculations on the light-position dependence of the shift current in a fermion chain. Sections~\ref{sec:discussion} and \ref{sec:summary} are devoted to discussions and summary, respectively. The nonequilibrium Green's function method for the Floquet theory and its numerical implementation are elaborated in \ref{sec:method} and \ref{sec:computation}, respectively. The detail of the quasiclassical theory for carrier concentration and the shift current is explained in \ref{sec:carrier_density}.

\section{Model and Method} \label{sec:model}

\begin{figure}[tbp]
  \centering
  \includegraphics[width=\linewidth]{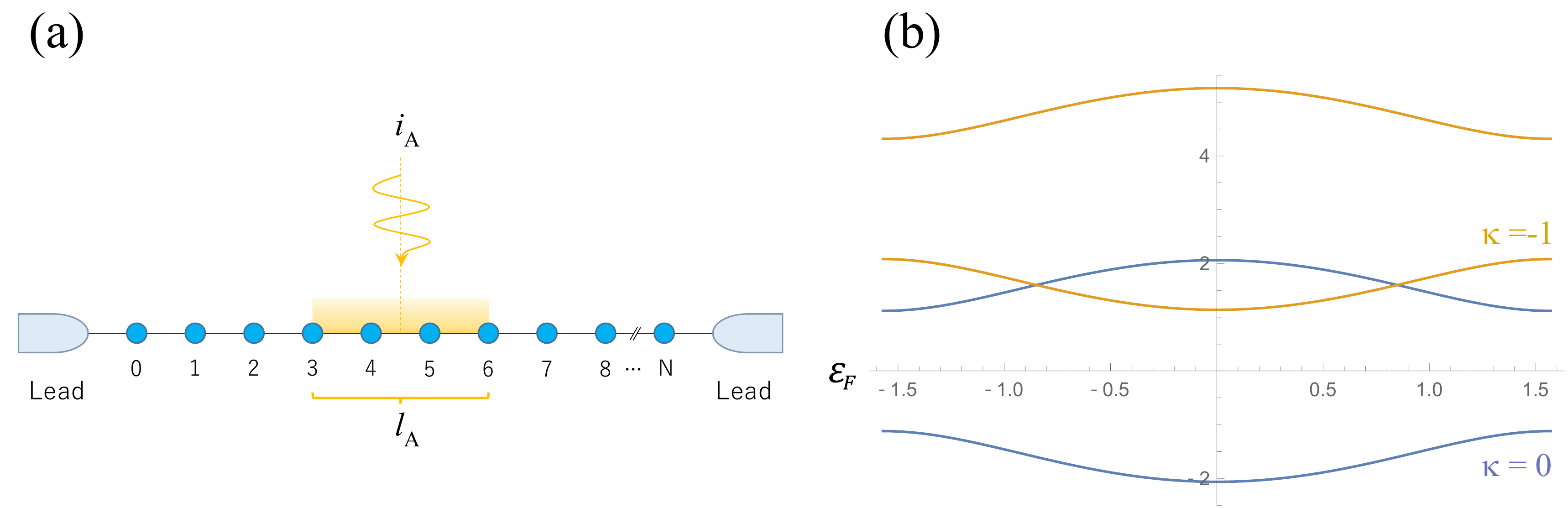}
  \caption{Schematic figure of the model. (a) an $N$ site fermion chain with bond alternation and staggered potential. The chain is coupled to the leads at each end. The incident light excites $l_A$ sites of the chain at $i_A$. (b) Floquet bands of the fermion chain in the bulk limit ($N\to\infty$). See the text for details.}
  \label{fig:model}
\end{figure}

To study the effect of local photo-excitation in noncentrosymmetric systems, we consider Rice-Mele model~\cite{Rice1982} coupled to leads (electrode) at each end and with monochromatic incident light [Fig.~\ref{fig:model}(a)]; the light is taken into account as the time-dependent electric field. The Hamiltonian for the chain is given by
\begin{eqnarray}
  H= H_0 + H_A,\label{eq:H}
\end{eqnarray}
where the first term is the Rice-Mele Hamiltonian,
\begin{eqnarray}
  H_0&=& -t\sum_i c_{i+1}^\dagger c_i + c_{i}^\dagger c_{i+1} - \frac{B}2\sum_i (-1)^i\left(c_{i+1}^\dagger c_i + c_{i}^\dagger c_{i+1} \right)\nonumber\\
  &&\qquad+ \frac{d}2\sum_i (-1)^i c_i^\dagger c_i, \label{eq:H0}
\end{eqnarray}
and the second term is the interaction with the incident light,
\begin{eqnarray}
  H_A&=& -itA\sin(\Omega \tau) \sum_i{}^\prime c_{i+1}^\dagger c_i - c_{i}^\dagger c_{i+1}
\nonumber\\
&& + \frac{iAB}2\sin(\Omega \tau)\sum_i{}^\prime (-1)^i\left(c_{i+1}^\dagger c_i - c_{i}^
\dagger c_{i+1}\right). \label{eq:HA}
\end{eqnarray}
Here, $c_i$ ($c_i^\dagger$) is the annihilation (creation) operator of the electron at the $i$th site; $t$ ($B$) is the uniform (staggered) hopping between the nearest-neighbor sites, and $d$ is the staggered potential. In Eq.~(\ref{eq:HA}), $\Omega$ is the frequency of the light and $A$ is the amplitude of vector field. Here, we only consider the leading order in $A$. The sum $\sum^\prime$ is over consecutive $l_A$ sites centered at $i_A$ [see Fig.~\ref{fig:model}(a)].

In periodically driven systems, there are no eigenstates as in the time-independent Hamiltonians. However, according to the Floquet theory~\cite{Floquet1883,Hanggi2005}, one can define similar eigenstates for the periodically driven Hamiltonian called Floquet states. In the Floquet theory, in addition to the bands that consists of the eigenstate of time-independent Hamiltonian, we consider the side bands of the original band with the energy shift of $-\kappa\Omega$ ($\kappa\in\mathbb Z$) [Fig.~\ref{fig:model}(b)]. When the chain is exposed to light, it gives the mixing term between the side bands with different $\kappa$; the leading order contribution from the light comes in as the mixing between two neighboring subbands (subbands with the indices $\kappa$ and $\kappa+1$). Therefore, to take into account of the leading order contributions from the light, in the subsequent calculations, we focus on the bands with indices $\kappa=0$ and $-1$ bands and ignore all others (\ref{sec:computation}). 

To study the nonequilibrium transport phenomena in the Rice-Mele model, the photocurrent from the local excitation is studied using a nonequilibrium Green's function theory~\cite{Meir1992,Rammer2007,Haug2008}. We generalized the method to the case of periodically driven systems and computed the shift current numerically. The general formalism and further details on the application of the method to the model in Eq.~(\ref{eq:H}) are elaborated in~\ref{sec:method}.

\section{Results: Shift Current in Noncentrosymmetric Crystals} \label{sec:results}

\subsection{Light-Position Dependence of Shift Current}

\begin{figure}[tbp]
  \centering
  \includegraphics[width=0.65\linewidth]{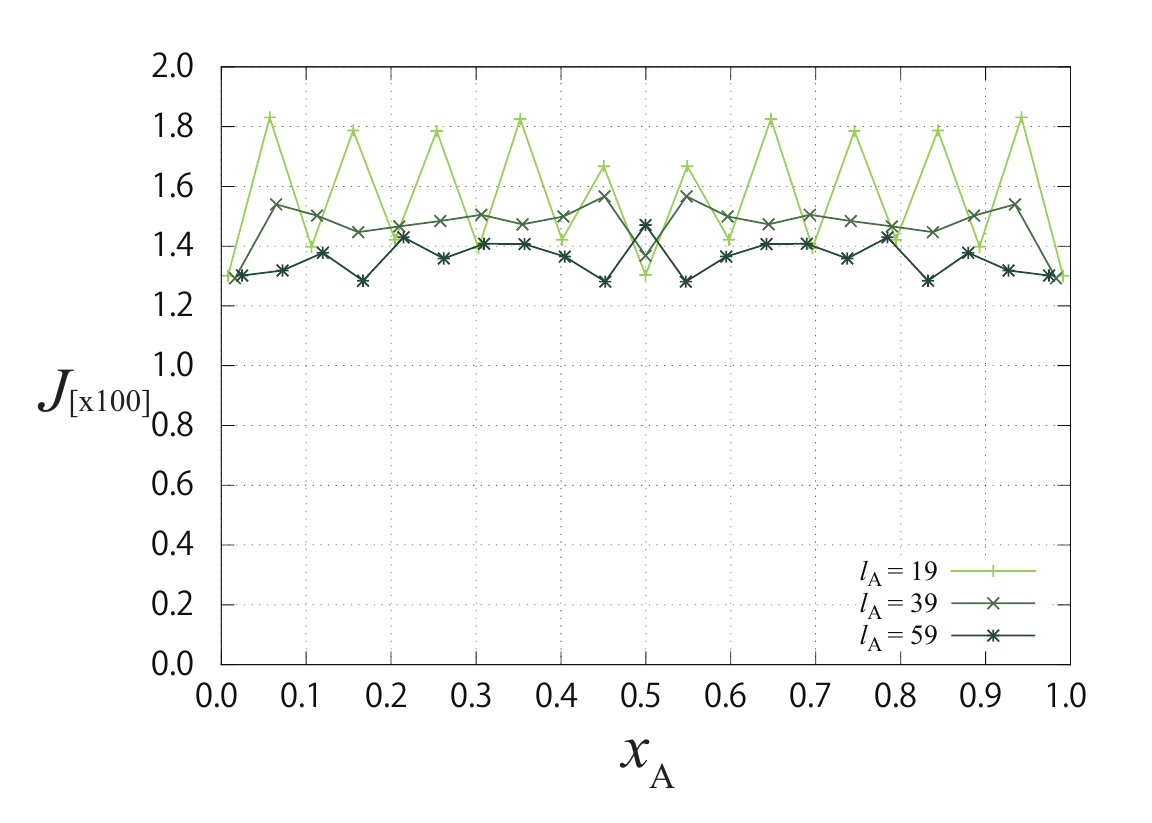}
  \caption{The result of the photocurrent in a $N=1200$ site fermion chain with bond alternation ($B=1$) and staggered potential ($d=1$). The figure shows the dependence of the photocurrent on the location of incident light ($x_A=i_A/N$) for different widths of light ($l_A$).}
\label{fig:bdchain}
\end{figure}

In this section, we study the shift current induced by the local excitation by light. Figure~\ref{fig:bdchain} shows the light-position dependence of the shift current by the local excitation at $x_A=i_A/N$ for $N=1200$. The three different lines correspond to different $l_A$, the number of bonds we shine the light onto [see fig.~\ref{fig:model}(a)]. The results show that the shift current by the local excitation is almost independent of $x_A$. Besides, the shift current is symmetric with respect to $x_A$, i.e., $J(x_A)\simeq J(1-x_A)$. For later convenience, we here define the ``symmetric'' and ``asymmetric'' components of the photocurrent by
\begin{equation}
  J_s(x_A)\equiv\frac{J(x_A)+J(1-x_A)}2
\end{equation}
and
\begin{equation}
  J_{as}(x_A)\equiv\frac{J(x_A)-J(1-x_A)}2,
\end{equation}
respectively. By calculating $J_s$ and $J_{as}$ from the data in Fig.~\ref{fig:bdchain}, we indeed find that $J_{as}\lesssim10^{-8}$, which is below the numerical precision. Hence, our results show that the symmetric part of the photocurrent is dominant in the Rice-Mele chain.

This is a contrasting feature from what is expected in the conventional mechanism in p-n junctions, by which the photocurrent is borne by charged quasi-particles. In p-n junctions, the photocurrent is essentially an interfacial phenomenon. At the interface (or surface), a potential gradient occurs due to the nonuniform structure. Due to this potential gradient, when an electron-hole pair is created by light, the electron and hole drift toward opposite directions, resulting in a net charge current. In the setup shown in Fig.~\ref{fig:model}(a), the junction of the chain and the lead plays the role of the interface. Therefore, in this mechanism, $J$ is expected to increase as we approach the edges. Besides, if the carriers are generated close to the edge, we expect that the excited carriers can escape to the electrodes easily, before being scattered back into the valence band. Hence, the current is expected to be larger at the ends of the chain than in the center.

Another important feature is the symmetry. Although there is no potential gradient in the bulk, in principle, a finite photocurrent can appear from the conventional mechanism when the position of the light is sufficiently close to one of the edges. This is a mechanism similar to p-n junctions where the potential gradient appears due to the interface. If this mechanism is dominant, for a centrosymmetric chain (or if the effect of broken inversion symmetry is negligible), the photocurrent from the local excitation is expected to be antisymmetric. On the contrary, if the photocurrent is induced by the bulk effect, e.g., anomalous photovoltaic effect, then the $x_A$ dependence of the photocurrent is expected to be independent of the position of light for a sufficiently long chain, with possibly some effect of the surface near the edges, i.e., they are (almost) symmetric. Therefore, we expect the light-position dependence of the locally excited photocurrent reflects the microscopic mechanism of the photocurrent; $J_{as}$ represents the contribution of the edges, e.g. the photocurrent by the conventional mechanism at the edge, while $J_{s}$ reflects the bulk effect such as the anomalous photovoltaic effects or shift current. 

These observations imply that the symmetric feature of the photocurrent with respect to $x_A$ reflects the unconventional photocurrent in the noncentrosymmetric chains, namely, the shift current. Therefore, in this paper, we call the symmetric photocurrent ``shift current''. Our result in Fig.~\ref{fig:bdchain} indicates that shift current is the dominant source of the photocurrent in Rice-Mele chain. To compare the result of Rice-Mele model with a model with no shift current, we also performed the calculation of $x_A$ dependence for a symmetric chain with $t=1$, $B=1.2$, and $d=0$ (not shown). In this model, however, we find no observable photocurrent within the numerical precision (only small $J_{as}$ that is below our numerical precision), consistent with the above discussion. 

\begin{figure}[tbp]
  \centering
  \includegraphics[width=0.65\linewidth]{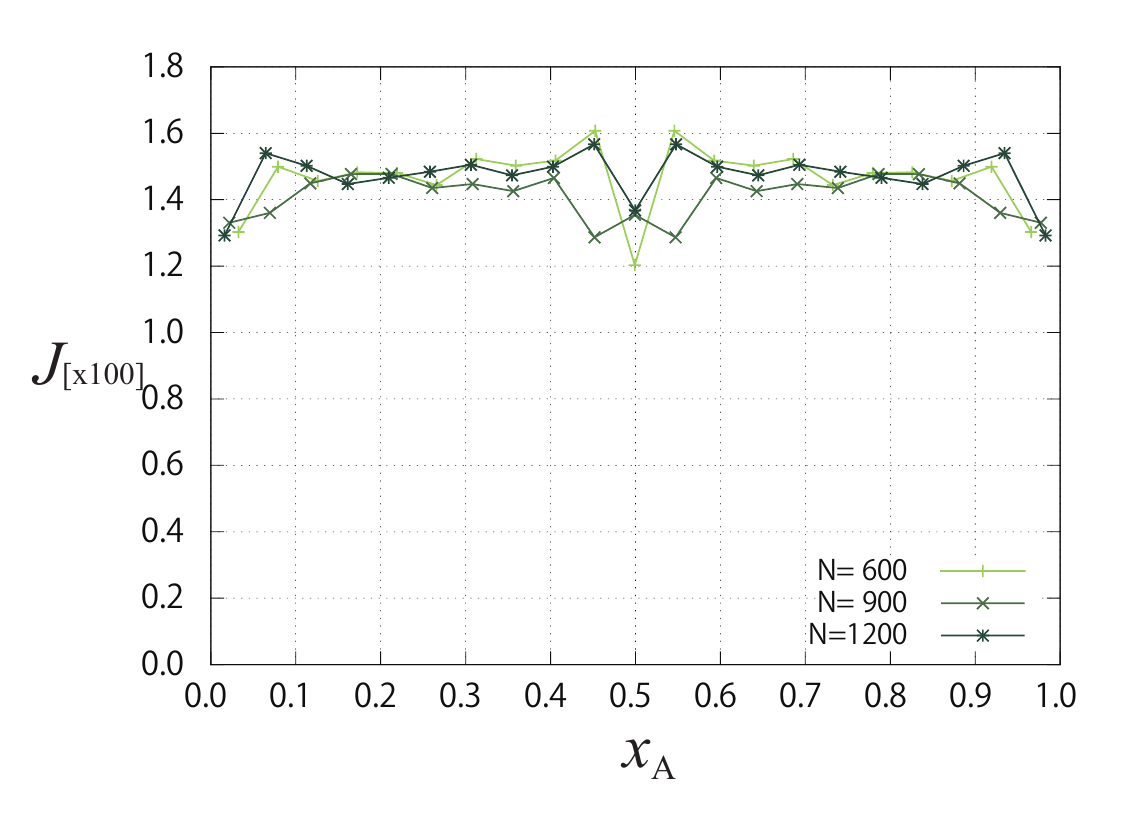}
  \caption{Light-position dependence of the photocurrent in the fermion chain with both the bond alternation ($B=1$) and the staggered potential ($d=1$). Each line shows the result for different length $N$.}
  \label{fig:chain_len}
\end{figure}

In the last, we discuss the effect of finite size effect. Fig.~\ref{fig:chain_len} shows the result of the asymmetric chain for $l_A=39$ with $N= 600$, $900$, and $1200$. The results for the different $N$ show similar magnitude of the shift current. They are also symmetric with respect to $x_A$. From this observation, we conclude that our data for $N=1200$ reflects $N\to\infty$ behavior in the ballistic limit.

\subsection{Light Width Dependence of Shift Current} \label{sec:results:lightwidth}

\begin{figure}[tbp]
  \centering
  \includegraphics[width=\linewidth]{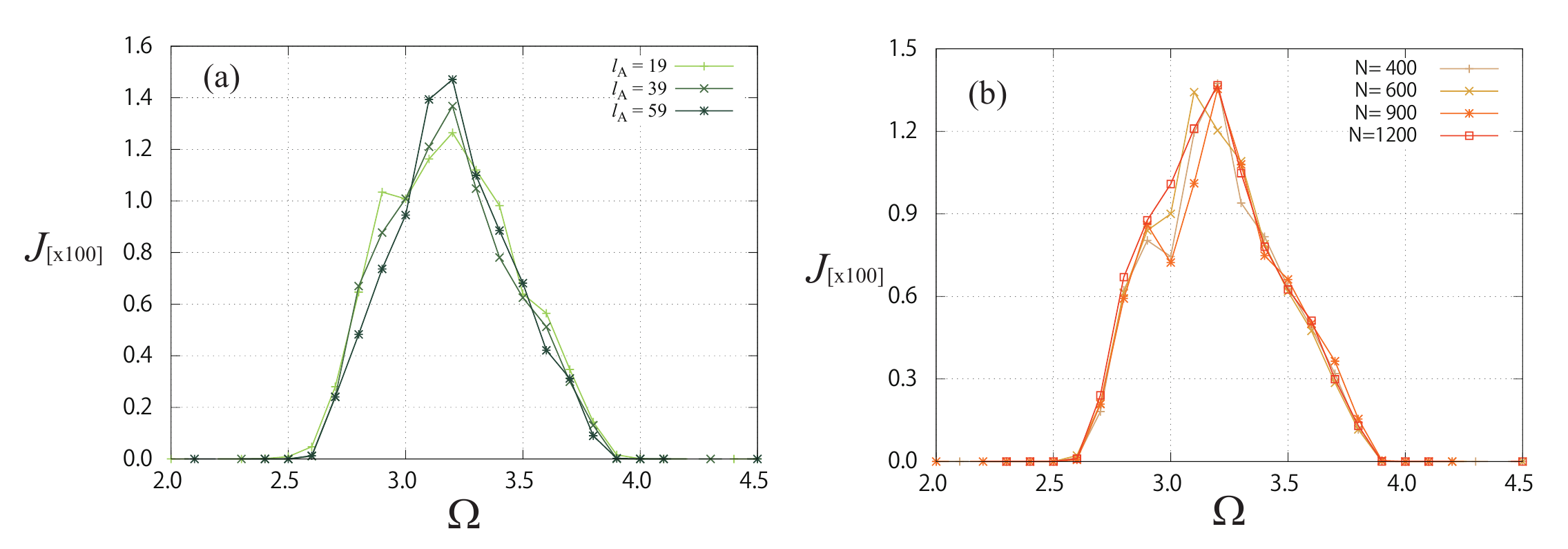}
  \caption{$\Omega$ dependence of the photocurrent in an noncentrosymmetric chain with $t=B=d=1$. The light is shined at the center of the chain ($x_A=i_A/N=0.5$). (a) photocurrent by local excitation with different widths of light $l_A=19$, $39$, and $59$. (b) photocurrent by the local excitation with different length $N$.}
  \label{fig:omega_dep}
\end{figure}

Another point to be noted in Fig.~\ref{fig:bdchain} is the $l_A$ dependence of the shift current. In Fig.~\ref{fig:bdchain}, the induced shift current appears to be roughly the same for all $l_A$. Naively, the photocurrent is expected to be proportional to the amount of the photon energy absorbed by the chain, therefore, proportional to $l_A$. Our results, however, show that there is nearly no dependence of the induced current on $l_A$. In our calculation, such behavior was observed for all range of $\Omega$. Figure~\ref{fig:omega_dep} shows the result of shift current excited by the local excitation at $x_A=0.5$. Figure~\ref{fig:omega_dep}(a) is the result for different $l_A$ in $N=1200$ noncentrosymmetric chain. A finite photocurrent is observed above $\Omega\sim2.6$ and form a triangular shape with a peak at $\Omega\sim3.2$. The lower limit $\Omega\sim2.6$ roughly corresponds to the band gap $\sqrt{\Delta^2+4B^2}=\sqrt5\sim2.24$, while the peak corresponds to an $\Omega$ which a subband of the conduction band perfectly overlap with that of the valence band, $\frac12(\sqrt{\Delta^2+16t^2}+\sqrt{\Delta^2+4B^2})=\frac12(\sqrt5+\sqrt{17})\sim3.17$. The results show almost no $l_A$ dependence for all range of $\Omega\in [2.5:4.0]$, where the shift current is observed.

Figure~\ref{fig:omega_dep}(b) shows the $N$ dependence of the shift current with $l_A=39$ and $A=0.2$ at $x_A=0.5$. For most values of $\Omega$, the results of shift current show almost no dependencies to $N$. On the other hand, some finite size effect is seen at $\Omega\sim3.1$. The $N$ sensitive behavior around $\Omega\sim3.1$ is likely to come from the sparse density of states in the center of the band. In the Floquet theory, the effect of light primarily appears around the crossing of two bands with different Floquet indices $\kappa$. In our model with $t=B=d=1$, it is around the crossing of the conduction band with $\kappa=0$ and valence band with $\kappa=-1$ (see Fig.~\ref{fig:model}); the two bands touch at their edge for $\Omega\sim2.24$, and crosses at the center of the bands for $\Omega\sim3.17$. As the density of states is relatively sparse at the center of the band than at the edges, the discreteness of the energy levels due to the finite length become more evident at the band center. However, the $N$ dependence at $\Omega\sim3.1$ also appears to be converged for the chains we used here ($N=1200$).

The result indicates that, for a sufficiently long chain, there seems to be almost no $l_A$ dependence. Furthermore, from the $\Omega$ dependence, it seems the $l_A$ independent behavior has less to do with the microscopic structure of the wave functions, e.g., Berry connection. According to the Floquet theory, as the effects of light have much smaller energy scale than other parameters, the effect of light on the electrons appears primarily around the band crossing point, where $\varepsilon_v(k)+\Omega=\varepsilon_c(k)$~\cite{Oka2009}. Here, $\varepsilon_v(k)$ is the energy eigenstate of the valence band with wavenumber $k$, and the $\varepsilon_c(k)$ being that of conduction band [see Fig.~\ref{fig:model}(b)]. As the crossing point changes by changing $\Omega$, if the behavior reflects the microscopic structure of the electronic states, e.g., Berry phase, then it is natural to expect $\Omega$ dependence. Therefore, it is unlikely that the $l_A$-independent behavior is related to the mechanism of the shift current.

\begin{figure}[tbp]
  \centering
  \includegraphics[width=0.5\linewidth]{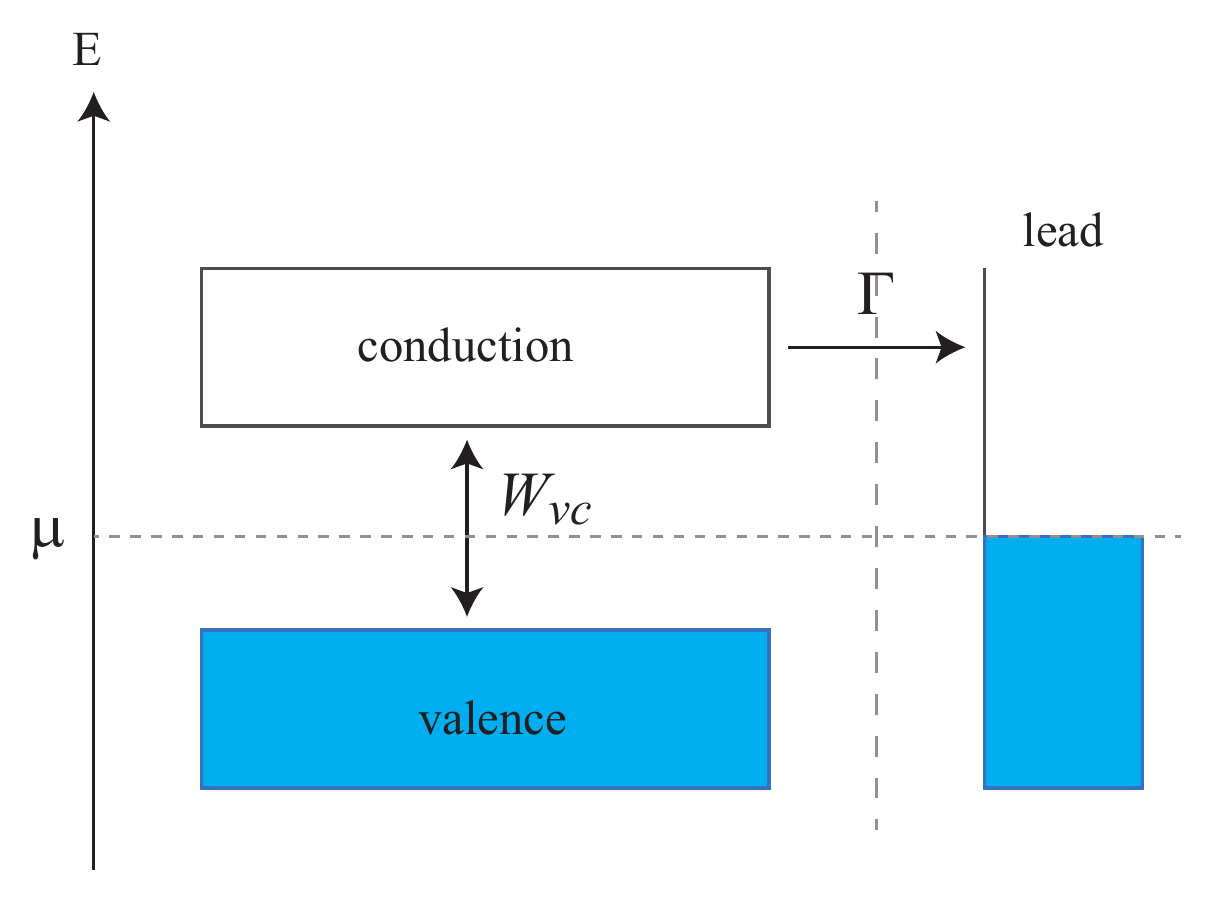}
  \caption{Schematic figure of charge transfer between the conduction and valence bands of a system irradiated with light and coupled to a lead. We assume that the lead is in its thermodynamical equilibruium state.}
  \label{fig:quasi_classic}
\end{figure}

From this observation, the $l_A$-independent behavior is likely to come from the fact that a large number of carriers are excited in the conduction band. To show this, we here consider a quasi-classical model schematically shown in Fig.~\ref{fig:quasi_classic}. The model consists of valence band, conduction band, and the lead(s) as shown in Fig.~\ref{fig:quasi_classic}. The carriers are transferred between the valence and conduction bands by transition rate $W_{vc}$ and between the conduction band and the lead by hybridization $\Gamma$. Further details on the model and the calculation are given in \ref{sec:carrier_density}.

From the general formalism of the Green's function theory, the charge current induced in the model is proportional to the product of $\Gamma$ and the change of electron distribution function from the equilibrium distribution, as shown in Eq.~\ref{eq:J}. Therefore, we here focus on the carrier density in the conduction (valence) band, $\delta n_c$ ($\delta n_v$), assuming that the current is proportional to $\delta n_c$ and $\delta n_v$. In our case, $W_{vc}$ comes from the irradiation of light. A perturbation theory based on Fermi's golden rule shows that this term is proportional to $l_A$.

In the steady state ($\delta\dot{n}_{c,v}=0$), the calculation based on this simple model gives
\begin{equation}
  \delta n_c = \frac{W_{vc} \rho_v}{\Gamma+W_{vc}(\rho_c+\rho_v)}.
\end{equation}
Here, we used charge neutrality condition,
\begin{equation}
  \rho_c\delta n_c = \rho_v\delta n_v.
\end{equation}
Similarly, we get
\begin{equation}
  \delta n_v = \frac{W_{vc} \rho_c}{\Gamma+W_{vc}(\rho_c+\rho_v)}.
\end{equation}
Therefore, for $W_{vc}\ll \Gamma$, the carrier densities reads $\delta n_{c,v}\propto W_{vc}\propto l_A$; the number of carriers and the photocurrent is proportional to $l_A$. On the other hand, if $W_{vc}\gg\Gamma$, $\delta n_c$ and $\delta n_v$ become (almost) independent of $W_{vc}$, i.e., it is independent of the width of excitation.

\begin{figure}[tbp]
  \centering
  \includegraphics[width=\linewidth]{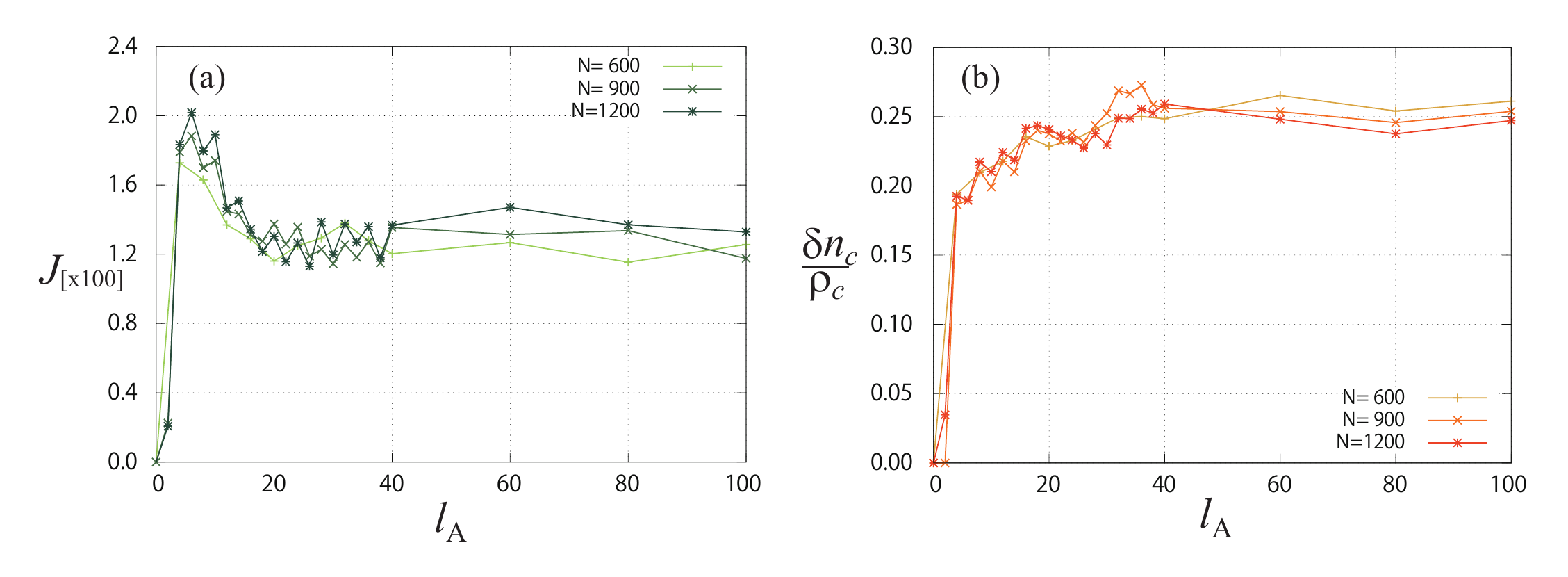}
  \caption{$l_A$ dependence of (a) the concentration of the excited electrons in the conduction band and (b) photocurrent in a noncentrosymmetric chain ($t=B=d=1$). The light is shined onto $l_A=39$ sites at the center of the chain ($x_A=0.5$. Different lines corresponds to the different length of the chain $N$.}
  \label{fig:carrier_dens}
\end{figure}

From the numerical calculations, it is difficult to directly give an estimate of $W_{vc}$. However, we can investigate to which limit the system is close to by calculating the density of excited carriers. If the system is in $W_{vc}\gg\Gamma$ limit, there exists a large density of carriers. On the other hand, the carrier density is about $\delta n_c\sim W_{vc} \rho_v/\Gamma$ for $W_{vc}\ll \Gamma$; the carrier density is much smaller than the density of state. In Fig.~\ref{fig:carrier_dens}, we calculated the carrier density in the conduction band with respect to the number of states, $\delta n_c/\rho_c$. The number of carriers is calculated by
\begin{eqnarray}
  \delta n_c = \frac1{2\pi N} \Im\left\{{\rm Tr}\int_0^\infty d\omega\, G^<_{00}(\omega)\right\},
\end{eqnarray}
and the number of the number of states by
\begin{eqnarray}
  \rho_c = \frac1{2\pi N} \Im\left\{{\rm Tr}\int_0^\infty d\omega\, \left[ G^A_{00}(\omega)- G^R_{00}(\omega)\right]\right\}.
\end{eqnarray}
Here, $G^<_{00}(\omega)$ is the $N\times N$ matrix of lesser Green's function and $G^{R,A}_{00}(\omega)$ is the retarded and the advanced Green's function with Floquet index $\kappa=0$. The $l_A$ dependence of $\delta n_c$ is plotted in Fig.~\ref{fig:carrier_dens}(a). Here, the calculation were done for $l_A=39$ and $x_A=0.5$ with $A=0.2$. The three curves were for the different length of chain, $N=600,900,1200$. For the noncentrosymmetric chain, the result shows a rapid increase of the carrier density below $l_A=20$ and a saturation to $\delta n_c/\rho_c\sim0.25$ above that. In coincidence with the saturation of the carrier density, the photocurrent also shows saturation above $l_A=20$. These features are consistent with the above mechanism for the $l_A$-independent behavior of the shift current.

\subsection{Current-Voltage Characteristic}

In this section, we study the current-voltage characteristic of the chains with shift currents by applying the external bias to the leads on the both sides. In our calculation, however, we find no current dependence on the external voltage within the numerical precision, if the bias is smaller than the charge gap (not shown). This is a natural consequence as the chain have no density of states at the Fermi level. As the carriers have higher energy than the bias, the effuluence of carriers to the leads is not affected by the change of electron distribution in the leads, which only takes place close to the Fermi level.

\section{Discussion} \label{sec:discussion}

Our results presented in Fig.~\ref{fig:bdchain} show that a large shift current can be induced in a ferroelectric chain by local excitation.  In the clean limit of a fermion chain, our result implies that the photocurrent in bulk photovoltaic materials can be robustly transported through the material for a long distance without decaying. This sheds light on to the potential realization of a highly efficient photovoltaic device using ferroelectric insulators~\cite{Choi2009,Young2012_1,Young2012_2,Grinberg2013,Nie2015,Shi2015,deQuilettes2015}.

An important point related to the robustness of the current is the effect of impurities. In experiments, the impurities often suppress electric current. In one-dimensional systems, the fermions are susceptible to Anderson localization. If the localization occurs, the current is likely to be suppressed at the scale of the localization length, i.e., no d.c. current is transported for the chains much longer than the localization length. Therefore, we expect the shift current to decay by the length scale of localization when a light is shined onto only a part of the chain.

In addition to the localization, another potential cause of the dissipation is the inelastic scattering by phonons. The inelastic scattering may contribute to the decay of shift current, limiting the distance the current is transmitted. Study on the effect of electron-phonon scattering to the shift current is left for future works.  

\section{Conclusion} \label{sec:summary}

In this paper, we theoretically studied the shift currents from local excitation by light. Considering a noncentrosymmetric insulating fermion chain, we studied the dependence of shift current on the position of excitation. By applying a non-equilibrium Green's function theory in \ref{sec:model}, we show that the local excitation in the noncentrosymmetric chain induces photocurrent that is almost independent of the position of excitation and is symmetric to the location of light ($x_A$), as shown in Fig.~\ref{fig:bdchain}. This is in contrast to the conventional mechanism, where the photocurrent is antisymmetric and maximized at the edge of the device, where the carriers can easily escape to the coupled leads. This difference in the position dependence of the photocurrent provides an experimental method to distinguish the unconventional photocurrent from that of the conventional ones, such as in the semiconductor junctions. Furthermore, this feature is advantageous for highly efficient optoelectronic devices.

\section*{Acknowledgements}

The authors thank M. Kawasaki, M. Nakamura, M. Sotome, Y. Tokura, and M. Ueda for fruitful discussions. This work was supported by JSPS Grant-in-Aid for Scientific Research (No. 16H06717, No. 24224009, No. 26103006, and No. 26287088), from MEXT, Japan, and ImPACT Program of Council for Science, Technology and Innovation (Cabinet office, Government of Japan).

\appendix
\section{Nonequilibrium Green's Methods}\label{sec:method}

To study the transport phenomena of the model in Eq.~\ref{eq:H} coupled to the leads, we use a nonequilibrium Green's function method for time-independent Hamiltonians~\cite{Meir1992,Haug2008}. In this section, we summarize the basic formulation we use to study the photocurrent. The details of the numerical implementation are given in \ref{sec:computation}.

\subsection{Keldysh-Floquet Theory for Systems Coupled to Leads}

For our calculation, we use an extension of non-equilibrium Green's function approach used widely to study non-equilibrium phenomena~\cite{Meir1992,Haug2008}; the theory is extended to Floquet space in order to study periodically driven systems~\cite{Hanggi2005,Haug2008}. In this section, we elaborate the general formalism and the derivation of Dyson's equation which we use to calculate the Green's function. We also elaborate the application to the model we study in the main text.

In the Floquet theory, the Green's functions have additional indices that corresponds to the Floquet states~\cite{Floquet1883,Eliasson1992},
\begin{eqnarray}
G_{\kappa\kappa^\prime}(\omega) &=& G_{\kappa-\kappa^\prime}(\omega+\frac{\kappa+\kappa^\prime}2\Omega),\label{eq:keldysh-wigner}
\end{eqnarray}
where
\begin{eqnarray}
\hspace{-5mm}G_n(\omega)= \frac1T\int_0^T dt_a \int dt_r G(t_a+\frac{t_r}2,t_a-\frac{t_r}2)\,e^{{\rm i} (n\Omega t_a+ \omega t_r)}
\end{eqnarray}
is the Wigner representation of the Green's function,
\begin{eqnarray}
G(t,t^\prime)=G(t+T,t^\prime+T),
\end{eqnarray}
with $T=2\pi/\Omega$ being the periodicity of the Hamiltonian in the time direction. Using this notation, the general form of Dyson's equation for the free particle systems is generally given by
\begin{eqnarray}
  (\omega+\kappa\Omega)\,&&G_{\kappa,\kappa'}(\omega) - \sum_{\kappa^{\prime\prime}}{\cal H}_{\kappa,\kappa''}^0\, G_{\kappa'',\kappa'}(\omega) = 1.\label{eq:eom_keldysh}
\end{eqnarray}
Here, 
\begin{eqnarray}
\mathcal H_{\kappa,\kappa'}^0=\frac1T\int_0^T dt H^0(t)\,e^{\mathrm i (\kappa-\kappa') \Omega t}
\end{eqnarray}
is the Fourier transform of the single particle Hamiltonian.

As the Hamiltonian, we here consider a free fermion system coupled to free fermion leads,
\begin{eqnarray}
  H(t)&=& H^0(\{c_i\},\{c_i^\dagger\},t) + \sum_{k,\alpha} \epsilon_{k\alpha} d_{k\alpha}^\dagger d_{k\alpha} + \sum_{k,\alpha,i} V_{k\alpha,i} d_{k\alpha}^\dagger c_i + {\rm h.c.}.
\end{eqnarray}
Here, $c_i$ ($c_i^\dagger$) is the annihilation (creation) operator for the $i$th state in the system and $d_{k\alpha}$ ($d_{k\alpha}^\dagger$) is the annihilation (creation) operator for $k$th eigenstate in the $\alpha$th lead. $\epsilon_{k\alpha}$ is the eigenenergy for the $k$ th state in the $\alpha$th lead, and $V_{k\alpha,i}$ is the hybridization between the $k\alpha$th state and $i$th state in the system.

For this type of models, the leads can be treated as a self-energy. The explicit form of self-energy can be deduced as follows. From Eq.~\ref{eq:eom_keldysh}, we get the relation
\begin{eqnarray}
  \left[G_{\kappa,\kappa'}\right]_{k\alpha,i}(\omega) = \sum_j \left[g_{\kappa,\kappa}\right]_{k\alpha,k\alpha}(\omega) V_{k\alpha,j} \left[G_{\kappa,\kappa'}\right]_{j,i}(\omega), \label{eq:dyson2}
\end{eqnarray}
where $g_{\kappa,\kappa'}(\omega)$ is the one-particle Green's function for $V_{k\alpha,i}=0$. By substituting Eq.~\ref{eq:dyson2} to Eq.~\ref{eq:eom_keldysh}, the Dyson equation for $\left[G_{\kappa,\kappa'}\right]_{i,j}(\omega)$ reads
\begin{eqnarray}
  (\omega+&&\kappa\Omega)\,G_{\kappa,\kappa'}(\omega) - \sum_{\kappa^{\prime\prime}}{\cal H}_{\kappa,\kappa''}^0\, G_{\kappa'',\kappa'}(\omega) - \sum_{\kappa^{\prime\prime}} \Sigma'_{\kappa,\kappa''}(\omega)\,G_{\kappa'',\kappa'}(\omega) = 1,\nonumber\\\label{eq:dyson_ren}
\end{eqnarray}
with
\begin{eqnarray}
  \left[\Sigma'_{\kappa,\kappa}\right]_{i,j}(\omega) = \sum_{k\alpha}V_{i,k\alpha} \left[g_{\kappa,\kappa}\right]_{k\alpha,k\alpha}(\omega) V_{k\alpha,j}. \label{eq:S_imp}
\end{eqnarray}
Not that, in Eq.~(\ref{eq:dyson_ren}), the Hilbert space is reduced to the degrees of freedoms in the chain, not the direct sum of chain and the leads.

In this paper, we particularly consider the case in which the leads are coupled only to the ends of the chain. From Eq.~(\ref{eq:S_imp}), the self-energy that comes from the leads, $\Sigma'_{\kappa,\kappa}(\omega)$, reads
\begin{eqnarray}
  \left[ \Sigma'_{\kappa,\kappa}(\omega) \right]_{i,j}&=&
  \delta_{i,0}\delta_{j,0}i\Gamma_R(\omega)\left( 
  \begin{array}{cc}
  -\frac12 & 2f_L(\omega+\kappa\Omega)-1 \\
  0 & \frac12
  \end{array}
  \right)\nonumber\\
&&\quad+\delta_{i,N-1}\delta_{j,N-1}i\Gamma_L(\omega)\left( 
  \begin{array}{cc}
  -\frac12 & 2f_R(\omega+\kappa\Omega)-1 \\
  0 & \frac12
  \end{array}
  \right),
  \label{eq:Sigma}
\end{eqnarray}
where $\alpha=L$ and $\alpha=R$ corresponds to left and right leads, respectively, and
\begin{eqnarray}
  \Gamma_\alpha = 2\pi \sum_k |V_{k\alpha,i_\alpha}|^2 \delta(\omega-\epsilon_{k\alpha}).
\end{eqnarray}
Here, $f_R(\omega)$ [$f_L(\omega)$] is the distribution function of electrons in the right (left) lead, and $i_\alpha$ is the site the lead is coupled to: $n_L=0$ and $n_R=N-1$, where $N$ is the number of sites in the fermion chain. In our calculation, we employed the wide-band approximation, namely, assuming $\Gamma_\alpha(\omega)$ to be a constant $$\Gamma_\alpha(\omega)=\Gamma_\alpha=\Gamma.$$

\subsection{Calculation of the Current}

By using a similar procedure we used to derive our Dyson's equation, the charge current that flows into/out of the system can also be calculated from the Green's function of the chain~\cite{Meir1992}. In our setup, the current between the $\alpha$th lead and the system is given by 
\begin{eqnarray}
  J_\alpha&=& i q_e \sum_{k,i} \int \frac{d\omega}{2\pi} \left\{ V_{k\alpha,i} \left[G^<_{\kappa,\kappa}\right]_{i,k\alpha}(\omega) - V_{i,k\alpha}\left[G^<_{\kappa,\kappa}\right]_{k\alpha,i}(\omega)\right\}, \label{eq:Jorig}
\end{eqnarray}
where $V_{i,k\alpha}=(V_{k\alpha,i})^\ast$. Here, $\kappa\in\mathbb Z$ is an arbitrary number. Following a similar procedure in Ref.~\cite{Meir1992} and by substituting $\left[G^<_{\kappa,\kappa}\right]_{i,k\alpha}(\omega)$ by Eq.~\ref{eq:dyson2}, Eq.~(\ref{eq:Jorig}) reads
\begin{eqnarray}
  J_\alpha= i q_e \int \frac{d\omega}{2\pi}\; {\rm tr}&&\left[ f_\alpha(\omega) \hat\Gamma_\alpha(\omega)\left\{\hat{G}^r_{\kappa\kappa}(\omega)-\hat{G}^a_{\kappa\kappa}(\omega)\right\}+\hat\Gamma_\alpha(\omega)\hat{G}^<_{\kappa\kappa}(\omega) \right].\nonumber\\ \label{eq:current}
\end{eqnarray}
Here, $\hat\Gamma_\alpha(\omega)$ is a square matrix with indices that corresponds to each state of the system,
\begin{eqnarray}
  \left[\hat\Gamma_\alpha\right]_{i,j}(\omega) = 2\pi \sum_{k\alpha} V_{j,k\alpha} V_{k\alpha,i} \delta(\omega-\epsilon_{k\alpha}),
\end{eqnarray}
and $\hat{G}^<_{\kappa\kappa'}(\omega)$ [$\hat{G}^r_{\kappa\kappa'}(\omega)$, $\hat{G}^a_{\kappa\kappa'}(\omega)$] is the matrix for lesser (retarded, advanced) Green's functions with $\kappa$ and $\kappa'$ indices for the Floquet states.

In the case of the model in Eq.~(\ref{eq:H}), $\hat\Gamma_\alpha(\omega)$ reads
\begin{eqnarray}
  \left[\hat\Gamma_\alpha\right]_{i,j}(\omega) = \delta_{i,i_\alpha}\delta_{j,i_\alpha}\Gamma_\alpha(\omega)
\end{eqnarray}
for $\alpha=L, R$. Hence, Eq.~(\ref{eq:current}) become
\begin{eqnarray}
  J_\alpha= i q_e \int \frac{d\omega}{2\pi}\; && f_\alpha(\omega) \Gamma_\alpha(\omega) \left\{\left[G^r_{\kappa\kappa}\right]_{i_\alpha,i_\alpha}(\omega)-\left[G^a_{\kappa\kappa}\right]_{i_\alpha,i_\alpha}(\omega)\right\}\nonumber\\
  &&\qquad+i q_e \int \frac{d\omega}{2\pi}\Gamma_\alpha(\omega)\left[G^<_{\kappa\kappa}\right]_{i_\alpha,i_\alpha}(\omega). \label{eq:current2}
\end{eqnarray}
Using these results, the current that flows through the system is given by~\cite{Meir1992}
\begin{eqnarray}
  J&=&J_R-J_L.
\end{eqnarray}
In the wide band approximation, Eq.~(\ref{eq:current2}) reads
\begin{eqnarray}
  J_a&=& -i\Gamma_a\int\frac{d\omega}{2\pi}\left\{G_{\kappa\kappa}^<(\omega)+f_a(\omega)\left[G^R_{\kappa\kappa}(\omega)-G^A_{\kappa\kappa}(\omega)\right]\right\}_{i_a,i_a},\\
  &=& -i\Gamma\int\frac{d\omega}{2\pi}\left\{G_{\kappa\kappa}^<(\omega)+f_a(\omega)\left[G^R_{\kappa\kappa}(\omega)-G^A_{\kappa\kappa}(\omega)\right]\right\}_{i_a,i_a}.
\label{eq:J}
\end{eqnarray}
In the second equation, we set $\Gamma_\alpha=\Gamma$.

\section{Numerical Calculations} \label{sec:computation}

For numerical calculation using the method explained in \ref{sec:method}, we introduce some approximation to implement the theory onto the computer. In addition, to accelerate the calculation, we implemented a matrix decomposition method. In this section, we illustrate the details on the numerical implementation of the nonequilibrium Green's function formalism.

\subsection{Truncation of the Dyson Equation}

The Green's function for our model in Eq.~(\ref{eq:H}) can be calculated from Eq.~(\ref{eq:dyson_ren}), by inverting the matrix
$$(\omega+\kappa\Omega)\hat1-{\cal H}^0-\Sigma',$$
where ${\cal H}^0$ is a matrix of ${\cal H}^0_{\kappa\kappa'}$ given by
\begin{eqnarray}
  {\cal H}^0=\left(
  \begin{array}{cccc}
    \ddots  & \vdots & \vdots &  \\
    \cdots  & {\cal H}^0_{\kappa\;\kappa} & {\cal H}^0_{\kappa\;\kappa+1} & \cdots \\
    \cdots  & {\cal H}^0_{\kappa+1\;\kappa} & {\cal H}^0_{\kappa+1\;\kappa+1} & \cdots \\
     & \vdots & \vdots & \ddots
  \end{array}
  \right),
\end{eqnarray}
and $\Sigma'$ is defined in the same manner with $\Sigma'_{\kappa\kappa'}$; $\hat1$ is the unit matrix. This matrix, however, is an infinite dimension matrix due to the Floquet indices. In our calculation, as we are interested in the leading order effect by the external light, we truncate the Floquet indices and leave the states that are close to the Fermi level, i.e., $\kappa=0$ and $-1$. Physically, this corresponds to considering only the leading order of scattering between the electrons in the valence band and that in the conduction band induced by light.

One point to be noted is that, due to this approximation, the periodicity of the Green's function with respect to $\omega$ is violated. To reduce the error that arises from the above approximation, for the calculation of shift current using Eq.~(\ref{eq:J}), we used $G_{00}(\omega)$ for $\omega\ge0$ and $G_{-1-1}(\omega)$ for $\omega<0$.

\subsection{Eigendecomposition of $G^{-1}(\omega)$} 

For the calculation of the Green's functions in long chains, an efficient algorithm based on the eigendecomposition of matrices is used. Here, we briefly review the algorithm we used in this calculation. Suppose we have a $\omega$-independent Hamiltonian $H$ and self-energy $\Sigma$. Then the inverse of the Green's function, $\omega\hat1-H-\Sigma$, is decomposed as
\begin{eqnarray}
  \omega\hat1-H-\Sigma = R(\omega-\Lambda)L^T,
\end{eqnarray}
where $R=(\vec\epsilon_1,\vec\epsilon_2,\cdots)$ and $L=(\vec{e}_1,\vec{e}_2,\cdots)$ are the $4N\times4N$ matrices of eigenvectors ($\epsilon_i$ and $e_i$ are right and left eigenvectors), and $\Lambda$ is a $4N\times4N$ diagonal matrix with eigenvalues as its diagonal components. Using this decomposition, the Green's function is given by
\begin{eqnarray}
  G(\omega)= R(\omega-\Lambda)^{-1}L^T.
\end{eqnarray}
Therefore, if the Hamiltonian and the self-energies are $\omega$ independent, then the arbitrary element of the Green's function can be calculated efficiently. 

In our formalism, in general, the self-energies are $\omega$ dependent. However, within the wide-band approximation and at zero temperature, the $\omega$ dependence of the self-energy appears as a step function, as shown in Eq.~(\ref{eq:Sigma}). Therefore, for a time-independent model, the above method is applicable by splitting the $\omega$ into two sectors, above and below the chemical potential. In the Floquet formalism, the number of sectors is given by $N_s+1$, where $N_s$ is the number of subbands. For the numerical integration in the calculation of the shift current using Eq.~(\ref{eq:J}), we find that the cutoff of $\Lambda_\omega^\pm=\pm4$ and discreteness of $\delta\omega=10^{-3}$ gives sufficient convergence.

\begin{figure}[tbp]
  \centering
  \includegraphics[width=0.65\linewidth]{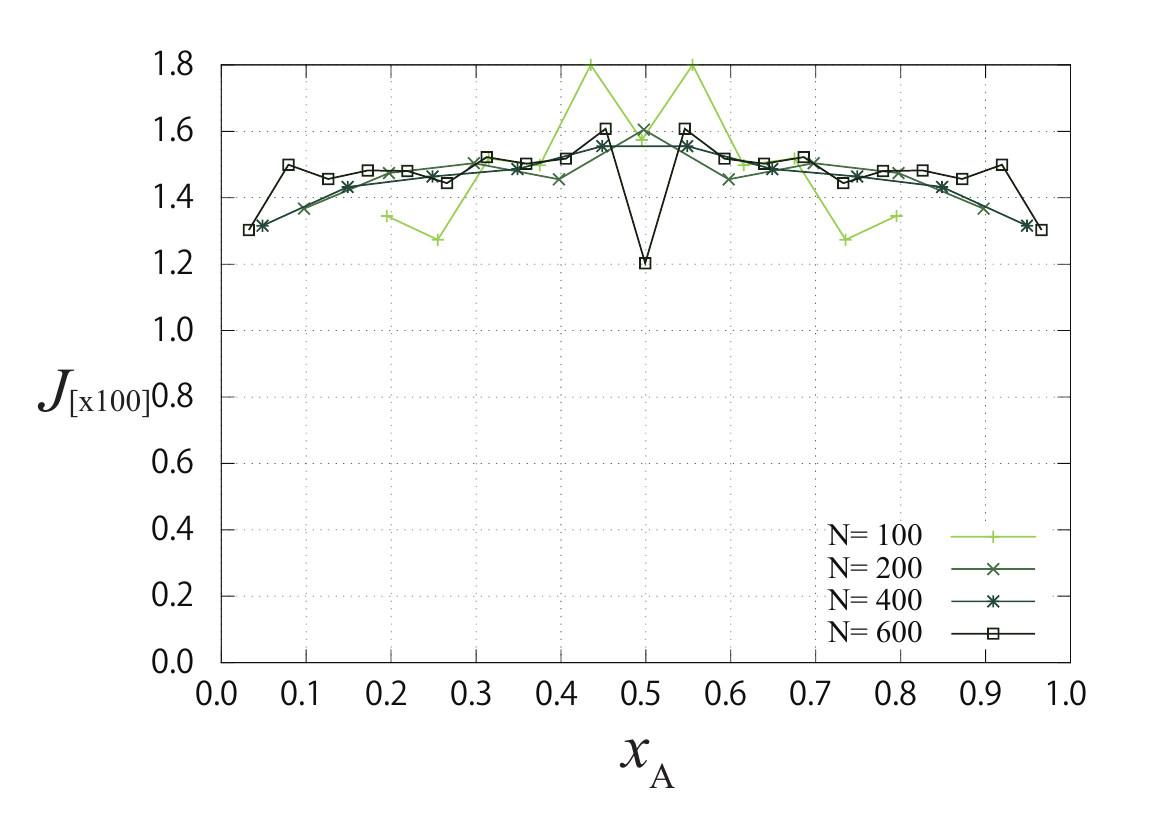}
  \caption{The finite size effect of photocurrent calculated with $l_A=39$ on a model with $B=1$ and $d=1$. The transverse axis is the location of incident light ($x_A=i_A/N$).}
  \label{fig:size_eff}
\end{figure}

In Fig.~\ref{fig:size_eff}, we show an example of the calculation using the approximation introduced above. The figure shows the results of the light-position dependence of shift current for $l_A=39$ with different length of chain (the results for larger sizes are given in Fig.~\ref{fig:chain_len}). The result for $N=100$ shows deviation from the results for larger $N$ with an arch-like structure. The results for $N\ge200$ show almost the same magnitude of the photocurrent and it is nearly independent of $x_A$. The oscillation of the result around $x_A=0.5$, on the other hand, remains robust for larger sizes. Presumably, this is related to the competition between the typical distance between the states in a band and the broadening of the pole of Green's function due to the leads. In general, the typical distance between the states is proportional to $1/N$; the numerical calculation converges when the typical distance become smaller than that of the broadening of the poles. In our setup, on the other hand, we only couple the edge sites to a lead. Hence, the poles of the Green's function become sharper with increasing size deep inside the bulk. Due to the competition between the two energy scales, the finite size effect around $x_A$ persists up to larger sizes. In the main text, we mostly use $N=1200$ for calculation.

\section{Quasiclassical Analysis on the Carrier Density} \label{sec:carrier_density}

This section elaborates the quasiclassical method discussed in Sec.~\ref{sec:results:lightwidth}. In the Green's function theory developed in \ref{sec:method}, the current given in Eq.~(\ref{eq:J}) is proportional to the change of the charge density from its ``equilibrium'' distribution, $f_\alpha(\omega)\rho_{i_\alpha}(\omega)$, where $\rho_{i_\alpha}(\omega)$ is the local density of states at site $i_\alpha$. Therefore, for the analysis on the $l_A$-independence, we focus on the carrier density in the conduction band. We consider a simple model that consists of a valence band, a conduction band, and the lead(s) as shown in Fig.~\ref{fig:quasi_classic}. The carriers are transferred between the valence and conduction bands by $W_{vc}$ and between the conduction band and the lead by $\Gamma$. The carrier concentration of the conduction band is given by solving a differential equation:
\begin{eqnarray}
  \rho_c \delta \dot{n}_c&=&-W_{vc}\rho_v\delta n_v\rho_c\delta n_c + W_{vc}\rho_v(1-\delta n_v)\rho_c(1-\delta n_c) - \Gamma \rho_c \delta n_c.
  \label{eq:phenom1}
\end{eqnarray}
Here, $\rho_{c}$ ($\rho_{v}$) is the density of states for conduction (valence) band, $\delta n_c$ ($\delta n_v$) is the concentration of the electrons (holes) in the conduction (valence) band. The first and the second term represents the transition of electrons between the conduction and valence bands (we discuss the detail in the following paragraphs), and the last term represents the effluence of carriers to the lead(s). The form of the last term can be guessed intuitively, or formally derived from the Keldysh theory, which the current that flows out of the chain is given by the product of hybridization and the modulation of electron density from the equilibrium distribution.

To give a qualitative understanding, we estimate $W_{vc}$ using a perturbative approach. From the Fermi's golden rule, for a perturbative Hamiltonian $H'(t)=H'(0)\sin(\Omega t)$, the transition amplitude from initial state $i$ to the final state $j$ reads
\begin{eqnarray}
  c_f(t)&=& \frac{i}{2\hbar} \langle f \left| H^\prime(0)\right| i\rangle \left[ \frac{\exp\left\{it(\omega_{fi}+\Omega)\right\}-1}{\omega_{fi}+\Omega} - \frac{\exp\left\{it(\omega_{fi}-\Omega)\right\}-1}{\omega_{fi}-\Omega} \right],\nonumber\\
\end{eqnarray}
where $\omega_{fi}\equiv\varepsilon_f-\varepsilon_i$ with $\varepsilon_i$ being the eigenenergy of $i$ th state. Here, we assumed the initial state $c_i(t)\sim 1$ and $c_f(t)\sim 0$ ($f\ne i$). In the $t\to \infty$ limit, the inside of the braces in this equation has two peaks in its absolute value, at $\omega=\pm \Omega$. We take the perturbative Hamiltonian to be the irradiation by light for electrons between sites $l_L$ and $l_R$ ($l_A=l_R-l_L$):
\begin{eqnarray}
  H_A(t) = -iA\sin(\Omega t) \sum_{n=l_L}^{l_R-1} \left(t-(-1)^n\frac{B}2\right)c_{n+1}^\dagger c_n - \left(t-(-1)^n\frac{B}2\right)c_n^\dagger c_{n+1}.\nonumber\\
  \label{eq:pert_fermi}
\end{eqnarray}
By the Fourier transformation of $c_n$'s, $H_A$ reads
\begin{eqnarray}
  H_A(t) &=& -iA\sin(\Omega t) \sum_{k,k'} I_{\frac{l_R-l_L-1}2}(k-k')\nonumber\\
  &&\times\left[ \left\{\left(t-\frac{B}2\right)e^{ik+i(k-k')l_L}-\left(t+\frac{B}2\right)e^{-ik'+i(k-k')(l_L+1)}\right\}c_{Bk}^\dagger c_{Ak'} \right.\nonumber\\
  &&\left.- \left\{\left(t-\frac{B}2\right)e^{-ik'+i(k-k')l_L}-\left(t+\frac{B}2\right)e^{ik+i(k-k')(l_L+1)}\right\}c_{Ak}^\dagger c_{Bk'} \right],\nonumber\\
  \label{eq:pert_fermi2}
\end{eqnarray}
with
\begin{eqnarray}
 I_n(dk) = \left\{
   \begin{array}{ll}
     \frac{n+1}N & \qquad(dk=0)\\
     \frac{1-e^{i2(n+1)dk}}{N(1-e^{i2dk})} & \qquad(dk\ne0)
   \end{array}
   \right..
\end{eqnarray}
Here, $c_{Ak}$ ($c_{Bk}$) are the annihilation operators for electrons on the even (odd) sites with wave number $k$. For simplicity, we assumed $l_L$ to be an even number site and $l_R$ to be odd. For a given $l_L$ and $l_R$, $I_n(dk)$ is given by a periodic function with periodicity $\pi$ (which corresponds to the size of the Brillouin zone), and with a peak of hight $(n+1)/N$ and the full width at half maximum $\sim \frac\pi{(n+1)}$. As the eigenstates for the bulk limit with wave number $k$ is given by a superposition of $c_{Bk}$ and $c_{Ak}$, the change in the wave number while exciting the electrons from the valence state to the conduction state is limited by the width of $I_n(dk)$.

In the $t\to \infty$ limit, the transition from the valence band electron with momentum $k$ to the conduction band electron with $k'$ occurs only between the $k$ and $k'$ with energies precisely separated by $\Omega$. The transition amplitude for a state in the valence band with $k$ to that in a conduction band with $k'$ is given by $\sim A^2|I_n(k-k')|^2\sim (l_A+1)^2/4N^2$ ($l_A=l_R-l_L$). Due to the constraint from energy conservation, the transition occurs only between the state close to the crossing between $\varepsilon_v+\Omega$ and $\varepsilon_c$ [see Fig.~\ref{fig:model}(b)]; the width of the wave number the transition takes place is given by $\sim \pi/n+1=2\pi/l_A+1$. Finally, the density of electron states per a width of $k$ is given by $N/2\pi$. From these facts, $W_{vc}$ is estimated to be
\begin{eqnarray}
  W_{vc}\sim A^2\left(\frac{l_A+1}{2N}\right)^2\frac\pi{l_A+1}\frac{N}{2\pi} = \frac{l_A+1}{8N}A^2 \sim\frac{l_AA^2}{N},
\end{eqnarray}
linearly proportional to $l_A$ and $A^2$.

In the steady state ($\delta\dot{n}_c=0$), the solution of Eq.~(\ref{eq:phenom1}) is given by
\begin{equation}
  \delta n_c = \frac{W_{vc} \rho_v}{\Gamma+W_{vc}(\rho_c+\rho_v)}.
\end{equation}
Here, we used charge neutrality condition,
\begin{equation}
  \rho_c\delta n_c = \rho_v\delta n_v.
\end{equation}
Similarly, we get
\begin{equation}
  \delta n_v = \frac{W_{vc} \rho_c}{\Gamma+W_{vc}(\rho_c+\rho_v)}.
\end{equation}

\end{document}